\documentclass[review]{elsarticle}
\usepackage{setspace}
\usepackage{graphicx}
\usepackage{amsmath}
\usepackage{float}
\usepackage{hyperref}
\usepackage{cleveref}

\title{Identifying Subcascades From The Primary Damage State Of Collision Cascades}

\author[addbarcv]{Utkarsh Bhardwaj}
\ead{butkarsh@barc.gov.in}

\author[addbarcv,addhbni]{Manoj Warrier}
\ead{manojwar@barc.gov.in}

\address[addbarcv]{Computational Analysis Division, BARC, Vizag, AP, India-530 012}
\address[addhbni]{Homi Bhabha National Institute,
Anushaktinagar, Mumbai, Maharashtra, India - 400 094}

\begin{document}
\begin{abstract}
The morphology of a collision cascade is an important aspect in understanding the formation of defects and their distribution. While the number of subcascades is an essential parameter to describe the cascade morphology, the methods to compute this parameter are limited. We present a method to compute the number of subcascades from the primary damage state of the collision cascade. Existing methods analyse peak damage state or the end of ballistic phase to compute the number of subcascades which is not always available in collision cascade databases. We use density based clustering algorithm from unsupervised machine learning domain to identify the subcascades from the primary damage state. To validate the results of our method we first carry out a parameter sensitivity study of the existing algorithms. The study shows that the results are sensitive to input parameters and the choice of the time-frame analyzed. On a database of 100 collision cascades in W, we show that the method we propose, which analyzes primary damage state to predict number of subcascades, is in good agreement with the existing method that works on the peak state.  We also show that the number of subcascades found with different parameters can be used to classify and group together the cascades that have similar time-evolution and fragmentation.  
\end{abstract}

  
 \maketitle

\section{Introduction} \label{sec:intro}

The collision cascades that are caused by high energy irradiation can have different cascade morphologies. One of the standard parameter to understand the cascade morphology is the number of subcascades. Subcascades are damage spots with high density of atomic displacements, separated from each other by less dense regions of damage \cite{heinisch1993structure, dierckx1987importance}. The collision cascades initiated by low energy primary knock-on atom (PKA) have a single damage spot. The PKA energy at which the average number of subcascades are greater than one is called the subcascade threshold energy and differs for different materials. At higher energies a collision cascade may get fragmented into smaller subcascades. A subcascade can be either well-separated from other subcascades or it may overlap with one or more subcascades \cite{antoshchenkova2015fragmentation}.

The cascade morphology and the formation of the subcascades affects the defect formation, defect sizes, defect morphologies and the spatial distribution of the defects \cite{de2018model, sand2017cascade}. Vacancies occupy the central region in a subcascade and interstitials form the surrounding in a roughly spherical fasion. This arrangement is very strictly found in low energy disconnected subcascades or low energy collision cascades. In case of connected subcascades, large SIA clusters are formed at the locations of the connections of the subcascades \cite{calder2010origin}. A cascade with more number of subcascades will also have a relatively higher fraction of smaller clusters of vacancies and interstitials
\cite{sand2017cascade}, except for the case of connected subcascades where large interstitial clusters get formed at the overlapping region. The damage from the disconnected subcascades caused by secondary knock-on atoms with a fraction of the energy of the PKA resemble a lower energy collision cascade. Above a certain energy threshold, cascade fragmentation becomes very probable and the damage regions do not show a linear increase in size and properties but rather they become a combination for lower energy cascades  \cite{Stoller2012293}. The time evolution of a cascade can be divided into three phases, initial recoil phase, peak displacement phase and the primary damage state at the end of the recombinations. The damage regions of all the phases have been shown to spatially correlate well with each other \cite{antoshchenkova2015fragmentation}.

Subcascades simulated with Binary Collision Approximation Monte-Carlo (BCA-MC) method have been extensively studied since decades using various methods such as fuzzy clustering \cite{hou1989fuzzy}, fractal method \cite{simeone2010cascade, satoh1991criterion, cheng1987cascade}, identifying vacancy rich regions \cite{satoh1991criterion} etc. A relatively recent BCA-MC study of subcascades uses a different method of decomposing the complete domain into cubes named as elementary cubes (ECs) and finding connected regions of cubes that are above an energy threshold to calculate the subcascades. A similar approach has been recently used for analysing the collision cascades in MD based on their peak damage state \cite{de2016subcascade}. The method requires the position and energy of all the atoms at the peak damage state of the collision cascade for analysis. The other important parameters of interest in collision cascades such as the number of defects, defect sizes, defect morphologies \cite{bhardwaj2021graph}, can all be calculated from the primary damage state. In databases of collision cascades such as the open CascadesDB database \cite{CascadesDB}, only the primary damage state is provided. Outputting the intermediate steps also increases the simulation output data size and impacts the efficiency due to high amount of data that is required to be written to disk. Moreover, it is the distribution of defects and damage regions at the primary damage state which are of interest from the point of view of initializing higher scale models of radiation damage with better spatial distribution. The higher scale models that incorporate spatial distribution such as the Kinetic Monte Carlo account for reaction rates more accurately compared to other models such as the Mean Field Rate Theory which lack spatial correlation \cite{stoller2008mean}. 

We present a method to calculate the number of subcascades using density based clustering algorithm (DBSCAN) applied on the vacancies found in the primary damage state. DBSCAN is a well establised unsupervised machine learning alogirthm \cite{mcinnes2017hdbscan} to classify dense regions into different clusters. The results of the DBSCAN algorithm with that from the EC method have been compared. We study the ECs method applied at the peak damage and extend it for better modelling of the subcascades count. We show that the nearby stages to peak in cascade evolution sometimes give a better idea about the number of subcascades. Moreover, the hyper-parameters in ECs method such as the threshold energy, can not be decided strictly. We propose a time averaged and parameter averaged way to deduce the number of subcascades and present the comparison of EC method with DBSCAN on a database of 100 collision cascades in W. We show that there is a good agreement between the two methods.

A classification of the collision cascades based on the subcascade volumes around the peak using unsupervised Machine Learning is presented. It is shown that the different cascade morphologies, such as the channelling cascades, connected subcascades, un-connected subcascades and single cascades form their own classes. These can be used to get important statistics for initialization in higher scale models to study the evolution of the defects. We explore the relationship of the classes with the properties of interest at primary damage state such as number of defects, maximum vacancy size, number of point defect clusters etc. A definite relationship of various classes and the properties of a collision cascade is observed and insights relating to morphology and the parameters is presented.

\section{Methods}
\label{sec:impl}

\subsection{Molecular Dynamics Simulation Dataset}
\label{sec:md}

We carry out subcascades analysis in the MD simulations of collision cascades in W. A 200 unit cells cubic bcc crystal of W is initially equillibrated at 0 bars pressure and 300K temperature. The collision cascades are initiated with a PKA energy of 150 keV. The simulations were carried out at an initial temperature of 300 Kelvin and evolved for 40ps. Electronic stopping was applied to atoms with energies above 10 eV \cite{hemani2020inclusion}. A PKA was selected at the center from the lattice atoms of the cubic simulation cell. The desired kinetic energy was given in a random initial direction. Periodic boundaries were used for each cascade. It was ensured that simulation cell size and PKA chosen were such that no atom reaches the boundaries except occasionally in the case of channelling. Temperature control at 300K was applied to all atoms within three atomic layers from the cell borders using a Berendsen thermostat \cite{doi:10.1063/1.448118}. MD simulations of high energy collision cascades with the Derlet potential \cite{PhysRevB.76.054107} stiffened by Bjorkas \cite{BJORKAS20093204} were carried out. The software used for MD simulations is LAMMPS \cite{thompson2022lammps}. The atomic coordinates are dumped every 500 time frames which is maximum of 0.5 picoseconds.

\subsection{Method for identifying subcascades from the primary damage state}
\label{sec:dbscan}

The algorithm to find the number of subcascades at the final primary damage state is based on the following facts:

\begin{itemize}
  \item The defect density in the three main stages of the collision cascade evolution viz. initial recoil phase, peak and final primary damage state, has strong spatial correlation \cite{antoshchenkova2015fragmentation}. 
  \item Subcascades can be defined as the regions of high damage separated by regions of lower damage \cite{de2016subcascade}. 
  \item The core of a subcascade has mostly vacancies surrounded by the interstitials \cite{calder2010origin}. 
  \item Density based clustering using the DBSCAN algorithm can group the high density regions into different clusters and ignore the low density regions as noise \cite{ester1996density} .
\end{itemize}

We identify high density regions of vacancies by using the DBSCAN clustering algorithm. The DBSCAN is a well established density based clustering algorithm used for unsupervised classification in Machine Learning. It can ignore the regions of low density vacancy defects that might appear in between two connected subcascades. It requires two input parameters viz. the maximum distance and the minimum number of points. The first parameter denotes the threshold distance between two points (here vacancies) beyond which they are not considered neighbors. The minimum number of points is another parameter that denotes the number of neighbours a point should have to be considered a core point. The core points grow the cluster by including the neighbouring points. The points in the cluster that are non-core are not checked further for neighbours to be included in the cluster.

The different values of maximum distance give the results at different level of details. We choose three different values of 20, 25 and 30 Angstroms, which is slightly higher values at primary damage than the values chosen for the length of effective cubes for the ECs algorithm at peak. Similarly, for minimum number of points we choose three different values, $2, 5$ and $10$, to get results at different level of details. The rounded-off value of the mean of subcascade counts found using different parameters is considered as the number of subcascades. The number of subcascades do not differ for different parameters in the case of well separated subcascades of substantial sizes. However, for the cases where the overlap may or may not appear depending on the level of detail, the results differ as we show later in the results section.

\subsection{Overview of the subcascade analysis at and around the peak}
\label{sec:overviewEC}

The subcascade analysis at the peak is done using the decomposition method as described in \cite{de2018model}. The method has been applied to the subcascade analysis
of BCA-MC \cite{de2016subcascade} and MD simulations \cite{de2021modelling}. The steps of the algorithm are given below:

\begin{enumerate}
\item Decompose the simulation box in equal sized cubic domains (ECs) of volume $V_c$ and length $l_c$.
\item Filter out the atoms that have kinetic energy below a threshold value $T_k$.
\item Find out the total energy of atoms in each EC.
\item Filter out the ECs that have the total kinetic energy lower than a threshol value $T_c$.
\item The volume of the cascade is the volume of the remaining ECs.
\item Find the peak frame as the frame at which the volume of the cascade is maximum.
\item At the peak frame, join the ECs that share an edge or corner together.
\item Each mutually connected group of ECs that is disconnected from others represents a different subcascade.
\end{enumerate}

The value of the hyper-parameters $l_c$, $T_k$ and $T_c$ have been suggested in earlier publications \cite{de2016subcascade, de2018model, de2021modelling}. We use values in the same range in our analysis.

The value of $l_c$ is chosen to be 1.5 nm. It has been shown that using different values of $l_c$ we can get slight differences in the total volume as well as number of subcascades \cite{de2018model}. The increase in the volume of the cascade with decrease in $l_c$ is due to the fractal nature of the collision cascade. With lower values of $l_c$ the number of subcascades increases separating out more and more subcascades due to smaller unconnected regions. For values between $1.0$ to $2.0$ nm the value of subcascades does not change much.

In previous studies \cite{de2016subcascade, de2021modelling}, the value of $T_k$ has been chosen from 0.1 to 0.4 eV. With higher value of the threshold, the subcascade count in most cases increases slightly because the cells falling in the connections of two overlapping subcascades get filtered out. However, a very high value may decrease the count by not accounting for smaller subcascades especially in the case of channelling.

$77 eV / {nm}^3$ has been suggested to be used as the value of $T_c$ for W. This is calculated based on the temperature needed to melt W \cite{de2016subcascade}. However, the exact calculation of the energy required to melt W in a non-equillibrium state might vary atleast from half of this value to $20\%$ higher \cite{de2018model}. The value used for $T_c$ varies from $27 eV / {nm}^3$ \cite{de2021modelling} to $77 eV / {nm}^3$ \cite{de2016subcascade}.

The number of subcascades may change depending on the value of the hyper-parameters $l_c$, $T_k$ and $T_c$ \cite{de2021modelling}. The exact values for these thresholds in the algorithms is ambiguous. Moreover, to understand the cascade morphology it is not sufficient to analyze just one frame at the peak \cite{de2021modelling}. In results section, we show cases where it is difficult to acertain the number of subcascades as ground truth. The values at different time-steps around the peak and with different hyper-parameters look reasonable and acceptable. Finally, for comparison with the results of the algorithm at primary damage state we use $T_k$ values of 0.1, 0.5 and 1.0 eV and $T_c$ values of $18$, $27$ and $54 eV / {nm}^3$. We take all the nine combinations of these two variables. The nine combinations of these values in a window of two time frames before and two time frames ahead from the peak make total of $9 * (2+2+1) = 45$. All these values are considered for the classification. The mean of these 45 values is taken as the fragmentation factor. The round-off of this fragmentation factor is the estimate of the number of subcascades. The mean of the inter-quartile range as the fragmentation factor can also be taken to remove the effect of outliers. However, in the present study we have taken the mean of all the values.

\subsection{Classification of cascade morphology}
\label{sec:classification}

Collision cascades are grouped into different classes based on their evolution and subcascade distribution. The number and volumes of subcascades at different time-frames around the peak are used as the feature vector for dimensionality reduction and unsupervised classification. The dimensionality reduction can help in visualizing and exploring the different classes having similar morphology. We use t-SNE \cite{van2008visualizing} for dimensionality reduction. It is a neighbour graph based dimensionality reduction technique that specifically focus on keeping a similar neighbourhood of the points in reduced dimensional space to the neighbourhood in higher dimensional space. For classification HDBSCAN \cite{mcinnes2017hdbscan} is used which is a density based classification algorithm very similar to DBSCAN.

\section{Results}
\label{sec:res}

The database analyzed contains 100 collision cascades simulated at 150 keV PKA energy for bcc W.

\subsection{Number of subcascades found at primary damage using the DBSCAN}
\label{sec:dbscanres1}

\begin{figure}[ht]
  \centerline{\includegraphics[width=.9\linewidth]{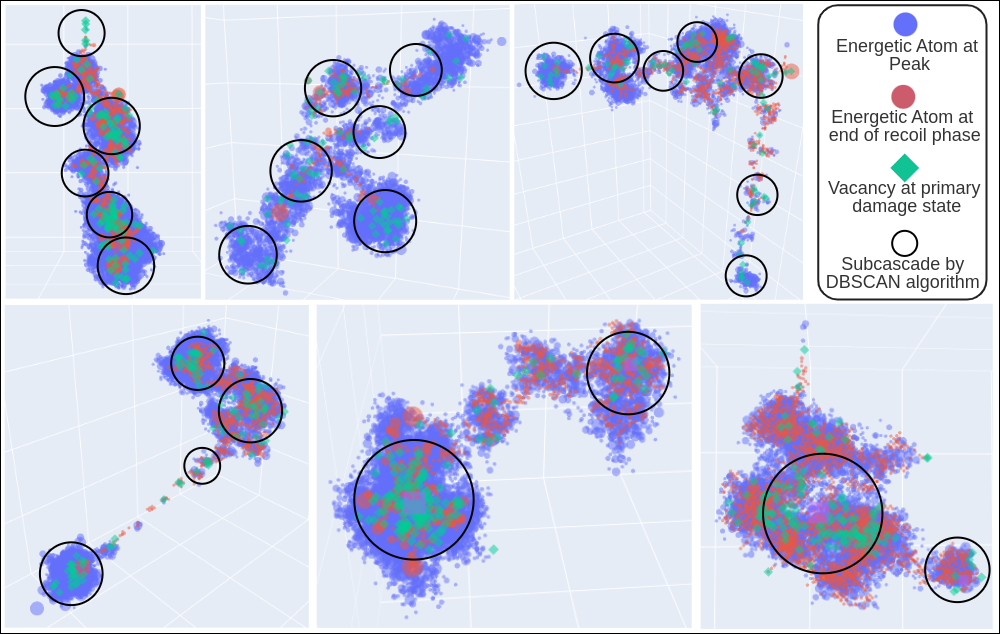}}
  \caption{\label{fig:fig-1}
    The damage regions at three different phases viz. at the end of recoil phase, peak damage and primary damage state of six different collision cascades. The subcascades found using DBSCAN algorithm at primary damage state overlap with the damage regions at all the phases.
  }
\end{figure}

\Cref{fig:fig-1} shows the subcascades found using the DBSCAN algorithm and the high energy atoms at the peak and the end of the recoil phase of the cascade. The marked subcascades can be seen as the regions of high vacancy core. The damage regions in all the three phases overlap.

\subsection{Comparison of EC method for different parameters and time frames around the peak}
\label{sec:ecres1}

\begin{figure}[ht]
  \centerline{\includegraphics[width=.9\linewidth]{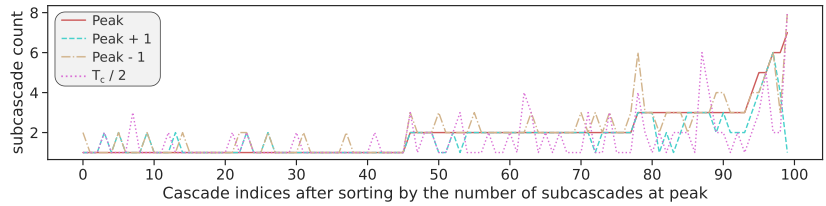}}
  \caption{\label{fig:fig-2}
    Number of subcascades found with different parameters. The variations are small especially for the lower subcascade counts.
  }
\end{figure}

\Cref{fig:fig-2} shows a comparison of the number of subcascades found for different cascades with different hyper-parameters and time-frames using ECs. The x-axis represents all the 100 cascades sorted by the number of subcascades found at the peak frame with $T_k = 0.1 eV$ and $T_c = 27 eV / {nm}^3$. The y-axis shows the number of subcascades for different parameters represented by the different colors and line-styles. We see that there are slight differences in the exact value of the number of subcascades for both the adjacent frames around the peak as well as if we decrease the value of $T_c$ by half. The reasons why the number of subcascades differ are listed below.

\begin{figure}[ht]
  \centerline{\includegraphics[width=.9\linewidth]{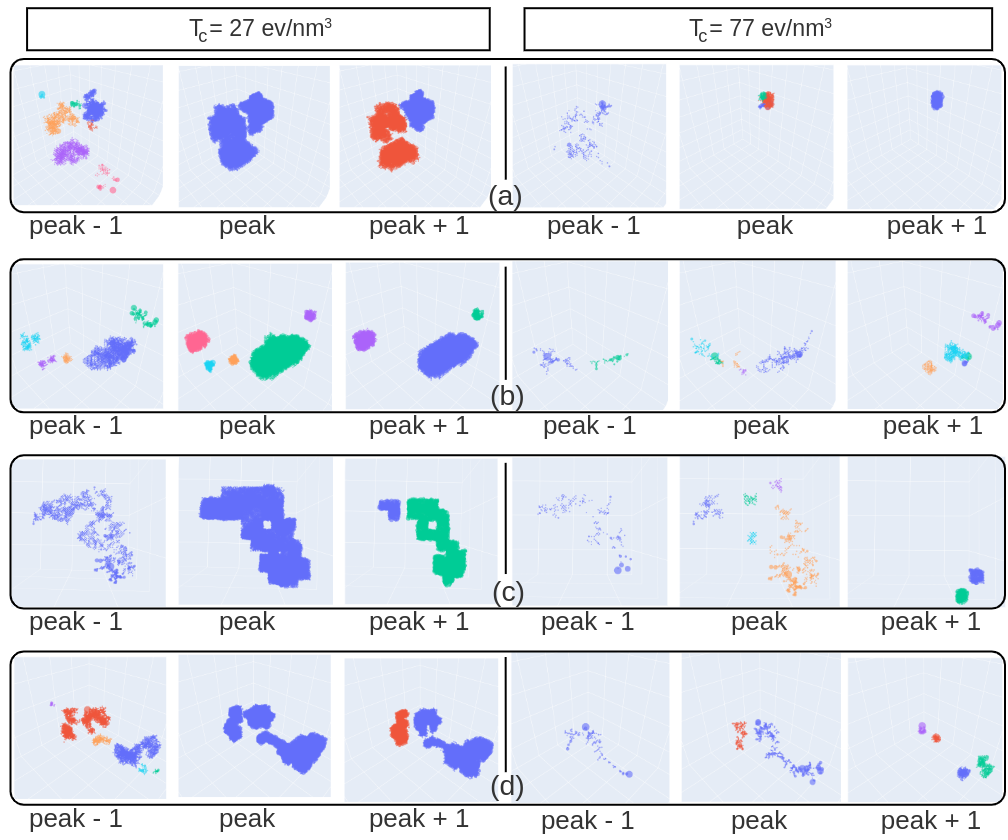}}
  \caption{\label{fig:fig-3}
Four example collision cascades showing subcascades at peak and its two adjacent frames for the two $T_c$ values of 27 eV/${nm}^3$ (left three plots) and 77 eV/${nm}^3$ (right three plots). The different subcascades identified with the EC method are coloured with different colours. It can be noted at peak that for lower $T_c$ value, the nearby subcascades merge into one while for higher $T_c$ value, the smaller damage regions might disappear. The adjacent frames are 500 simulation steps apart which is less than 0.5 ps.
  }
\end{figure}

\begin{enumerate}
  \item The slightly connected damage regions that may appear as one overlapping region at peak may become disconnected if we look at the adjacent frames especially when the $T_c$ is less as shown in \Cref{fig:fig-3}. This may also happen as we raise the value of $T_c$ at peak.
  \item Some of the low energy small disconnected subcascades may fade away before the peak volume of whole cascade is reached based on the volume of bigger subcascades. This is especially true when $T_c$ value for is high as shown in \Cref{fig:fig-3}.
  \item In case of channelling when a lot of cascades are created far from each other the peak of one region of damage might not be the peak of the other region, in these cases the concept of peak is weak and might give high or low number of subcascades. This is the reason why we see more disagreements in the region where subcascade count is particularly high due to channelling of the knock-on atoms.
  \item The connected subcascades may show overlap when threshold energies ($T_c$, $T_k$) are low while if we increase the threshold energies the less dense regions of connectivity may disappear \Cref{fig:fig-3}.
  \item Some of the smaller subcascades that are often less dense might also disappear if the threshold energies are increased \Cref{fig:fig-3}.
\end{enumerate}

Inspite of all these differences the overall distribution of the number of subcascades is not significantly different for small changes in the hyperparameters. However, if we take any one value as the ground truth for the number of subcascades, it may not be suggestive of the morphology and evolution of the cascade e.g. a cascade with a value of one subcascade at the peak might actually be a single spherical cascade or an overlapping connected pair of subcascades or it might have a small off-shooting subcascade that subsides before the peak is reached. It has been shown that subcascade count itself is not a good indicator of cascade morphology \cite{de2021modelling}. For this reason we take the feature vector of all the values found with different parameters and around the peak frame as the basis for classifying cascade morphology for which the results are shown in \Cref{sec:class}.

\subsection{Comparison of EC method and DBSCAN}
\label{sec:ecdbscan}

\begin{figure}[ht]
  \centerline{\includegraphics[width=.9\linewidth]{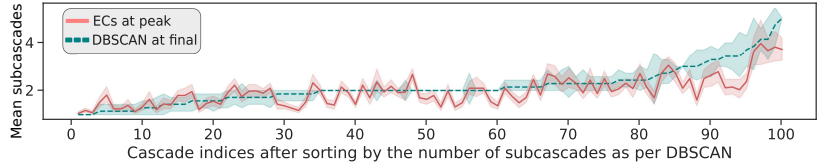}}
  \caption{\label{fig:fig-4}
    Number of subcascades found with the EC method and DBSCAN method with different parameters. The lines show the mean and the shaded region show the 95\% confidence region. The mean values are similar and the shaded regions generally overlap.
  }
\end{figure}

\Cref{fig:fig-4} shows the comparison of the subcascade counts found around the peak and the subcascade counts found at the primary damage state with different parameters. The mean value is shown with the line while the shaded region shows the 95\% confidence region.The values are shown for all the hundred collision cascades in the dataset that are sorted by the value found using DBSCAN. We see that the two methods almost always have overlapping shaded region. The average difference between the mean subcacade count for each cascade found using the two different methods is 0.37. For 95\% of the cascades the difference is below 1.0 and for 75\% for the data the difference is below 0.5. The maximum difference between the mean value of subcascades found using the two methods is 1.4. The cases where the differences are more show greater variability due to change in parameters as well.

\begin{figure}[ht]
  \centerline{\includegraphics[width=.9\linewidth]{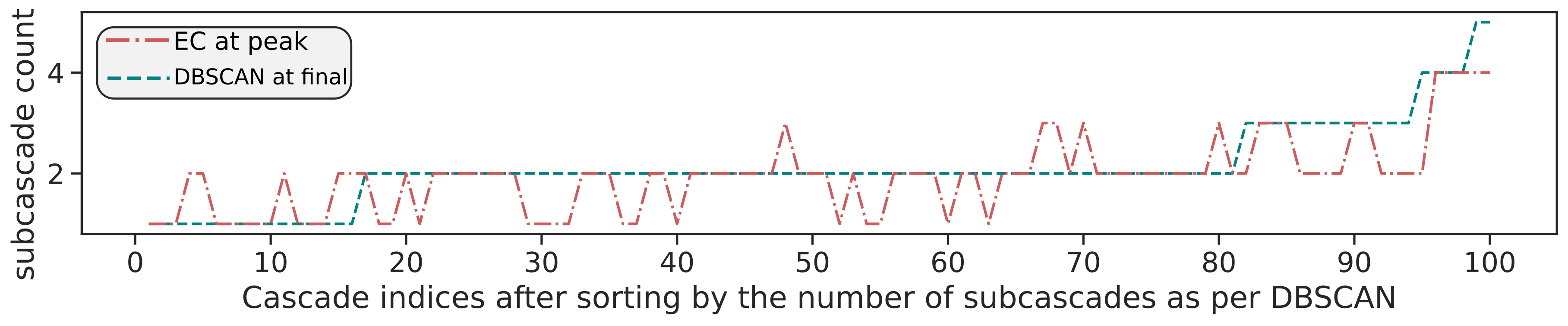}}
  \caption{\label{fig:fig-5}
    Number of subcascades found with the EC method and DBSCAN method after rounding up the mean value for different parameters. The methods show good agreement and the variations are within the same range as found with different parameters for the same method.
  }
\end{figure}

The rounded up mean value of the number of subcascades found using EC method and using DBSCAN are shown in \Cref{fig:fig-5}. The difference in most of the cases is limited to one however at higher number of subcascades the difference can be slightly more. This variation is similar to the variation observed between the subcascade counts found using different parameters at the peak (\Cref{fig:fig-2}). For 64\% of the cascades the values match exactly while in 35\% the values differ by one. Only one cascade shows a differnce of two. The small differences that we see arise in the cases where the ambiguity is inherent as described earlier.

\subsection{Classification of cascade morphologies}
\label{sec:class}

\begin{figure}[ht]
  \centerline{\includegraphics[width=.5\linewidth]{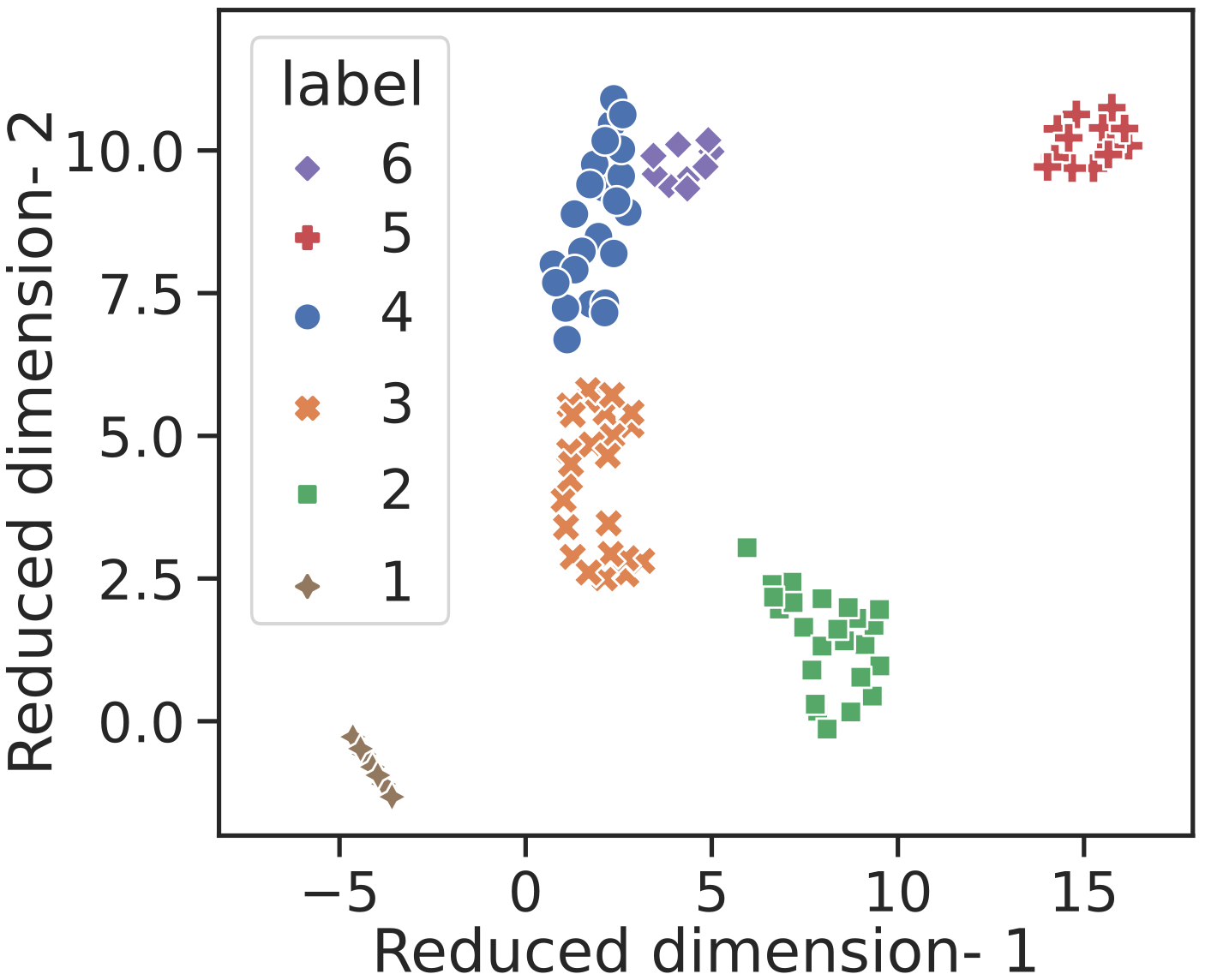}}
  \caption{\label{fig:fig-6}
    The similarity relationship and classes of all the collision cascade morphologies. Each point represents a collision cascade. The points representing collision cascades with similar morphologies are placed close to each other using t-SNE dimensionality reduction applied on the feature vector based on subcascade volumes. Different colours represent different classes of morphologies found using HDBSCAN density based clustering algorithm.
  }
  \end{figure}

The single value of subcascade count does not fully define the cascade morphology. We use subcascade count and subcascade volumes found using the different parameters and around the peak for representing the cascade morphology. To view the different classes of morphologies, we use the feature vector that consists of these values and apply the dimensionality reduction algorithm t-SNE to reduce it to two dimensions for plotting. \Cref{fig:fig-6} shows a plot where each point represents a collision ccascade. The two axes represent the first and second coordinate values found using the dimensionality reduction.

\begin{figure}[ht]
  \centerline{\includegraphics[width=.9\linewidth]{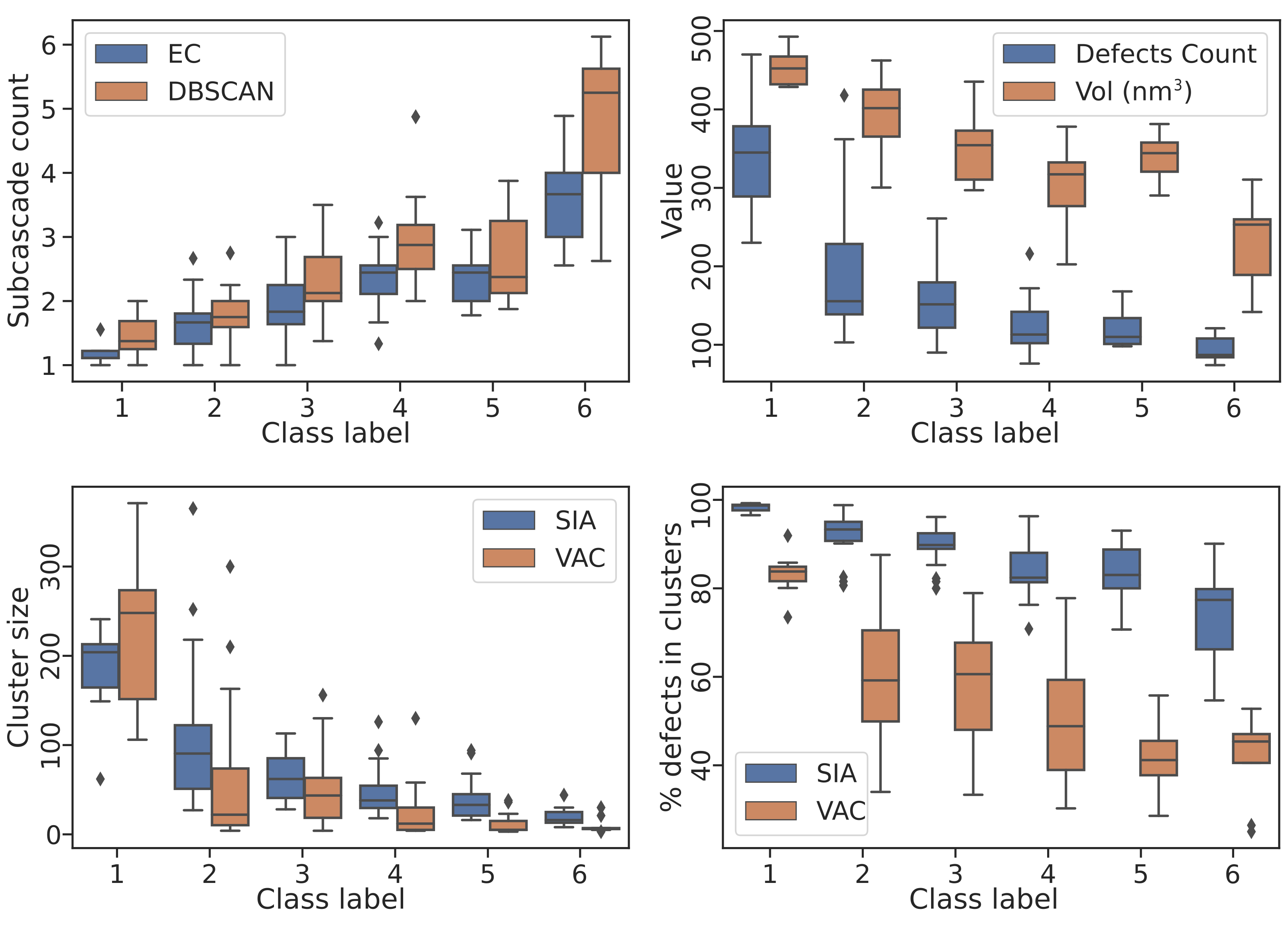}}
  \caption{\label{fig:fig-7}
    Distribution of various cascade properties in different morphological classes.
  }
\end{figure}

  We then classify the different dense regions that signify prevalent morphologies into classes using HDBSCAN. We obtained six classes on the current dataset. \Cref{fig:fig-7} shows distribution of various properties in different morphological classes. 

\begin{figure}[ht]
  \centerline{\includegraphics[width=.9\linewidth]{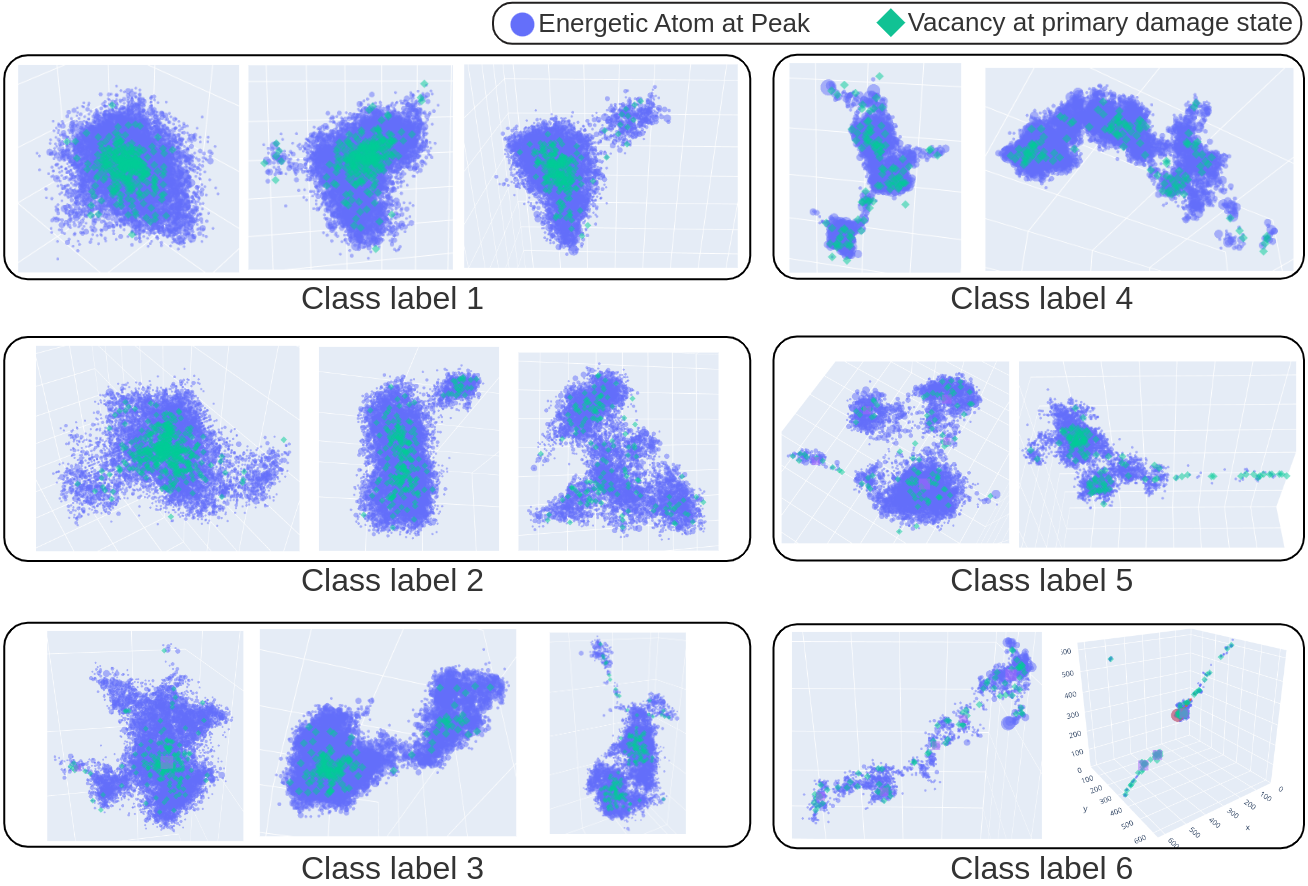}}
  \caption{\label{fig:fig-8}
    Sample cascade morphologies for each class labels.
  }
\end{figure}

The class label-1 represent collision cascades with only a primary knock-on and the mean number of subcascades slowly increase with class label. The fractional value of subcascade counts after taking the mean of subcascade counts using the different parameters is used for plotting. The collision cascades from label-1 to label-4 have subcascade counts between 1 and 3 but all other properties such as defects count, cluster sizes, percentage of defects in clusters show a decreasing trend. The properties of the label-5 only show typical trends in cluster sizes and to some degree in defects count as well. In \Cref{fig:fig-6} label-5 appears separate from all the other morphologies. \Cref{fig:fig-8} shows typical cascade morphologies for each label to show what the labels mean qualitatively.

The class label-1 represents collision cascades with only a single major subcascade which may also have a small slightly connected off-shoot. This small subcascade would rarely be counted. In 2nd class the single subcascade starts to appear more fragmented and the off shooting subcascade becomes more prominient in some cases. In 3rd and 4th cases the fragmentation further increases and more distinguished subcascades also appear. In label-5 the subcascades are further apart and there is sometimes a small energy channelling in addition to the bigger subcascades similar to label-4. Also, the vacancies seem to be smaller and more sparsely distributed. The label-6 has the cascades that have long range channelling sometimes piercing through the box boundaries which is large enough for all other cases.

\section{Conclusion \label{discuss}}

The method developed in this study identifies the number of subcascades using just the primary damage state. This opens up the door to analyze vast databases such as the CascadesDB database of collision cascades that only have the primary damage data. The number of subcascades can be estimated from the final frame just like all the other important parameters such as the number of defects formed, defect morphologies etc. The application of well-established techniques from machine learning such as DBSCAN and t-SNE have enabled us to provide an efficient and easy to understand method for the estimation of the number of subcascades and the classification of cascade morphologies.

We have also shown a classification of the cascade morphologies based on the subcascades volumes throughout evolution and with different levels of details. The classes show regular trends with the other parameters of interest which provide useful insights. The classes also show qualitative morphological characteristics which are not captured in the subcascade count alone. The morphologies such as channeling, single cascades which may or may not have nearby small subcascades etc. fall in different classes. The classification can be used in an exploratory way or may be used to classify new cascades into one of the known categories. In future more feature vectors or representations of collision cascades can be tried for such a categorization and can be used to study morphological correlations with defect morphologies etc. This information can be used to initialize the collision cascades in a higher scale model like KMC.

The dataset used in this work has a single material and uses the same interatomic potential. The distribution of collision cascade morphologies may change depending on the material and the interatomic potential. The comparative study of cascades morphologies across different interatomic potentials is out of the scope of the paper.

\section*{Data availability}
The raw data required to reproduce these findings are partially available to download from the open database https://cascadesdb.org/. The code required to reproduce these findings is available to download from \\ https://github.com/haptork/csaransh/ as part of the open source Csaransh software \cite{bhardwajcsaransh}.

\bibliographystyle{elsarticle-num} 
\bibliography{content}

\end{document}